# Robust Video Watermarking using Multi-Band Wavelet Transform

Jamal HUSSEIN[1] and Aree MOHAMMED[2]

[1] Computer Science Department, University of Sulaimani
Sulaimani, Iraq

[2] Computer Science Department, University of Sulaimani
Sulaimani, Iraq

**Abstract**

This paper addresses copyright protection as a major security demand in digital marketplaces. Two watermarking techniques are proposed and compared for compressed and uncompressed video with the intention to show the advantages and the possible weaknesses in the schemes working in the frequency domain and in the spatial domain. In this paper a robust video watermarking method is presented. This method embeds data to the specific bands in the wavelet domain using motion estimation approach. The algorithm uses the HL and LH bands to add the watermark where the motion in these bands does not affect the quality of extracted watermark if the video is subjected to different types of malicious attacks. Watermark is embedded in an additive way using random Gaussian distribution in video sequences. The method is tested on different types of video (compressed DVD quality movie and uncompressed digital camera movie). The proposed watermarking method in frequency domain has strong robustness against some attacks such as frame dropping, frame filtering and lossy compression. The experimental results indicate that the similarity measure before and after certain attacks is very close to each other in frequency domain in comparison to the spatial domain.

***Keywords:*** *Copyright Protection; Frequency Watermarking; Vector Estimation; Mulit-Band Wavelet; Quality Measurement; Similarity Measurement;*

## 1. Introduction

Video watermarking refers to embedding watermarks in a video sequence in order to protect the video from illegal copying and identify manipulations. A variety of robust and fragile video watermarking methods have been proposed to solve the illegal copying and proof of ownership problems as well as to identify manipulations [1]. The methods can be divided into techniques that work on compressed or uncompressed data [2].
Various types of watermarking schemes have been proposed for different applications. For copyright-related applications, the embedded watermark is expected to be immune to various kinds of malicious and non-malicious manipulations to some extent, provided that the manipulated content is still valuable in terms of commercial significance or acceptable in terms of perceptual quality. Therefore, watermarking schemes for copyright-related applications are typically robust [3].

The watermarking techniques have been applied either in the spatial domain [4,5] or in the frequency domain using (Fourier, DCT, DWT, Fractal, etc) transforms [6,7].

In [12], a blind, imperceptible and robust video watermarking technique is proposed. This algorithm is based on cascading two mathematical transforms; the Discrete Wavelet Transform (DWT) and the Singular Value Decomposition (SVD).

This work focuses on the implementation and evaluation of robust watermarking technologies for color video with the intention to embed watermarks in every encoded video frame which is selected from motion estimation prediction. Also a new algorithm of color video watermarking applied in the frequency domain is proposed. This algorithm uses the predictive motion estimation [8] to find the best matched blocks in order to embed the proposed watermark (random Gaussian distribution) between frames. This motion information consists of motion vectors and prediction errors for each individual block. The former estimates the motion on a pel-by-pel basis, whereas the latter predicts the motion on a block-by-block basis [9]. The general flowchart diagram of the proposed work is shown in the Fig. 1. The necessary steps to embed the watermark into an input video data for the copy right protection purpose are as follows:
1. Extract loaded color video into frames.
2. Apply block matching motion estimation techniques on the subsequent frames.
3. Select only those frames that have sufficient number of motion blocks which is compatible with the watermark size.

4. From the selected frames use a given threshold to select the best blocks during the matching process.
5. Perform the wavelet transformation on the selected best blocks.
6. Embed a random Gaussian distribution as a proposed watermark into the selected blocks (Apply only to the HL and LH wavelet bands).
7. Extract the embedded watermark.
8. Apply some attacks on the watermarked frames in the video.
9. Evaluate the conducted results using PSNR for embedding and similarity for extracting process before and after attacks.

Note: the terms and definitions used in this paper are described in the appendix.

In the second section, we describe an existing technique for spatial video watermarking and adapt this algorithm to design a frequency based video watermarking. The test results and all relevant discussions appear in section III.

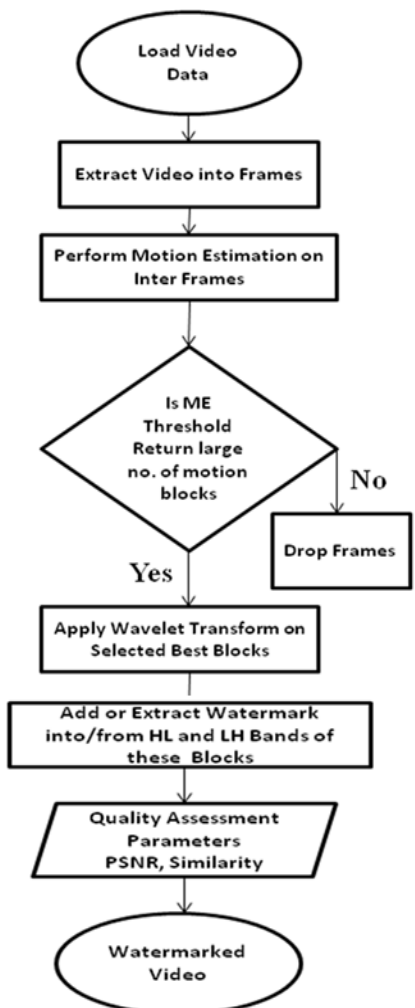

Fig. 1  Video watermarking flowchart diagram

The tests mainly based on motion estimation, adaptive quantization and malicious attacks. Finally, in section IV, we assess our achievements so far, and provide an overview of further work.

## 2. Video Watermarking Scheme

2.1 Spatial Domain

The proposed color video watermarking scheme in spatial domain is implemented through the following steps:

A. *Convert Video Color Space*

YCbCr refers to the color resolution of digital component video signals, which is based on sampling rates. In order to compress bandwidth, Cb and Cr are sampled at a lower rate than Y, which is technically known as "chroma subsampling." This means that some color information in the video signal is being discarded, but not brightness (luma) information.

For these reasons the proposed watermarking is added only to the Y component.

**Y = 0.2989 * R + 0.5866 * G + 0.1145 * B**
**Cb = -0.1687 * R - 0.3312 * G + 0.5 * B**
**Cr = 0.5 * R - 0.4183 * G - 0.0816 * B**

B. *Motion Estimation*

In this paper, the method used for predicting blocks motion vectors (motion estimation) is modified one at a time search algorithm (MOTS). The method is a modified version of the OTS standard method which is a very simple but effective algorithm; it makes horizontal and vertical scanning separately. During the horizontal scanning stage, the point of minimum distortion on the horizontal axis is being searched until it is found the horizontal scan stops. Then, starting from the stopping point, the minimum distortion in vertical direction is found. The modification is based on utilizing the existing inter-block correlation between the motion vectors of the adjacent blocks. MOTS use the predetermined motion vectors of the previous blocks (i.e., top, left, and top block to calculate the initial motion vector of the tested block). The main advantage of this modification is determining motion vectors with minimum prediction error, which is lower than that obtained by using traditional OTS [8]. The initial motion vectors (dx, dy) used in MOTS method are calculated according to the following equations:

$$\Delta x = \frac{1}{3}[x(i-1, j-1) + x(i, j-1) + x(i-1, j)] \quad (1)$$

$$\Delta y = \frac{1}{3}[y(i-1, j-1) + y(i, j-1) + y(i-1, j)] \quad (2)$$

For simplicity, the block sizes of integer power of 2 are preferred. In this work, the size of blocks was taken 8x8

pixels. For each frame to be watermarked, in the first step it will uniformly be divided into non-overlapped blocks.

### C. Block Selection Criteria

After the motion estimation is performed on the successive frames in a tested video, the best matching blocks are found. According to the size of the watermark (32*32), a specific number of blocks are needed. In the proposed scheme the nearest blocks from the center are chosen as a criterion. Fig. 2 illustrates the best chosen blocks among all motion blocks.

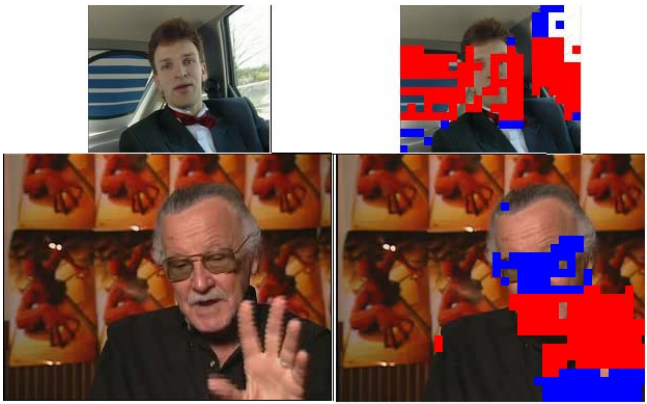

Fig. 2 Left: reference frame Right: all blocks (in blue), selected best blocks (in red)

### D. Watermark Generation

Given a source of uniform pseudo-random numbers, the Box-Muller transform in a polar form can be used to transform uniformly distributed random variables to a new set of random variables with a Gaussian or normal distribution [10]. In this paper, the generated watermark is represented as 1024 bytes or a matrix of (32*32).

### E. Watermark Embedding

At the first the proposed scheme performs the watermarking algorithm in the spatial domain. The spatial-domain directly modifies the intensities or color values of some selected pixels. The blocks are selected as shown in the section C.

Since the watermark size is (1024 bytes), we need 1024/(m*m) blocks to embed the watermark data in each frame, where m*m is the block size. For example if m = 8 (block size is 8*8) then 16 blocks are needed to embed the watermark.

The watermark should also be divided into blocks of the same size as the block's frame. Each block in the watermark added to a block from the frame by adding Y component of the pixel in watermark's block to the Y component in the corresponding pixel of the frame's block using the following equation.

$$I_w(x, y) = I(x, y) + \alpha W(x, y) \quad (3)$$

Where
$I$: the original frame data
$I_w$: the watermarked frame data
$W$: the watermark data
$\alpha$: the scaling factor
$x, y$: 0……m-1 where m is the block size.

This operation is repeated for every selected frame in the video.

As a measure of distortions introduced by watermarking process, the visual quality of the watermarked data is required to be as high as possible. Visual quality means that the degradation of the data due to the watermarking operation should be imperceptible. The Peak Signal to Noise Ratio PSNR is used as visual quality measurement.

$$\text{PSNR(dB)} = 10\log_{10}(m^2 \frac{\max(I(x,y))^2}{\sum(I(x,y) - I_w(x,y))^2}) \quad (4)$$

### F. Watermark Extraction

The watermarked frame and the scaling factor that used in embedding process, the watermark $W^*$ can be extracted from the original using the following equation:

$$W^* = \frac{I_w^*(x,y) - I(x,y)}{\alpha} \quad (5)$$

Where
$I$: the original frame
$I_w^*$: the possibly altered frame
$W^*$: the extracted watermark
$\alpha$: the scaling factor
x, y: 0..m-1 where m is the block size

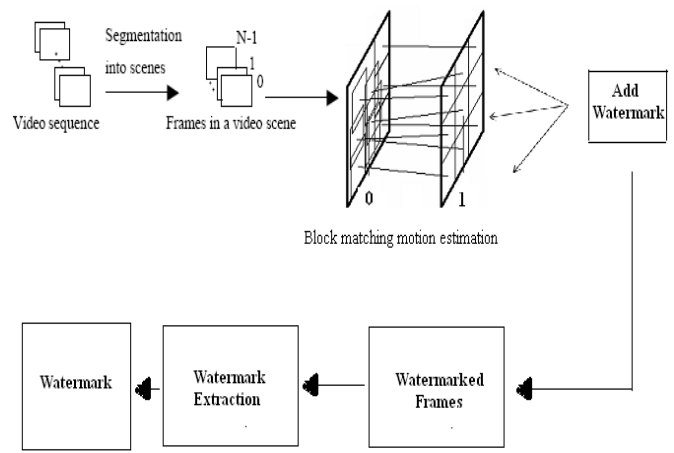

Fig. 3 Watermark Embedding and Extraction process in spatial domain

The insertion and extracting processes of the proposed watermark in the spatial domain can be summarized in fig. 3.

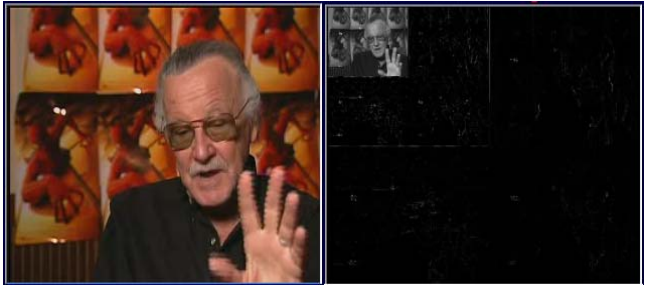

Fig. 4   Frame with 2 level wavelet decompositions

### G.   Quality Measurements

The extracted watermark W* is compared to the originally embedded watermark W using the similarity measure.

$$\delta = \frac{W^* \cdot W}{\| W^* \| \cdot \| W \|} \qquad (6)$$

The similarity $\delta$ varies in the interval [-1, 1]; the value in the interval (0, 1] indicates the extracted sequence W* matching the embedded sequence W and therefore one can conclude that the frame has been watermarked with W.

### 2.2 Frequency Domain

The video watermarking scheme in frequency domain follows the same steps as explained in the spatial domain except the way that the watermark is embedded or extracted to/from the wavelet blocks (HL and LH) bands. The reasons behind the embedding of the watermark into the HL and LH bands are:

1. LL band consists of a large amount of energy in the signal. Therefore, if there is an abrupt motion in the video frames, the inserted watermark cannot be robustly extracted when it is threatened by attacks.
2. HH band consists only of some details information and it is very fragile to embed watermark in it.

In this research, the filter used to apply wavelet transform is a non reversible biorthogonal transform (9/7 Tap) [11]. This transform can only be used for lossy coding. The FWT is applied on a frame by transforming the rows (as the first stage), and then the columns of the frame (as the second stage). It yields two-dimensional decomposition (four-channel decomposition). The wavelet transform compacts most of the image energy into the LL subband which consists of a few coefficients in comparison with the whole number of wavelet coefficients (which is equal to the total number of pixels). Unlike conventional transforms, wavelet decomposition produces a family of hierarchically organized decompositions. The selection of a suitable level of the hierarchy depends on the signal nature and experience. In this work, the wavelet decomposition level is 2.

For the second level of wavelet transform, since the watermark size is (1024 bytes), we need 1024/ (m/2+m/2) blocks to embed the watermark data in each frame, where m*m is the block size. For example if m = 8 (block size is 8*8) then 128 blocks are needed to embed the watermark.

In fig. 4 a reference frame of a video with two levels of wavelet decomposition is illustrated.

The insertion and extracting processes of the proposed watermark in the frequency domain (added only to HL and LH subbands) can be summarized in fig. 5. The main advantage of using the frequency domain in our video watermarking scheme is the resistance against some attacks. This evaluation is done through the similarity measure of a watermark before and after attack.

## 3.   Test Results Evaluation

The adopted test strategy was based on determining the effects of the involved parameters on the performance parameters (Similarity and PSNR) as follows:

1. Number of wavelet pass (taken as constant = 2).
2. Block size (taken as constant = 8).
3. Watermark size (taken as constant = 512 or 1024 bytes)
4. Threshold of motion estimation to return more than 128 motion blocks according to the watermark size (taken as constant = 4).
5. Scaling factor α has not a direct effect in frequency domain due to the small values of wavelet coefficients (taken as constant = 0.1).
6. If any frame contains less than 128 motion blocks, they consider as a dropped frames.
7. Similarity between original and extracted watermark before and after attacks (taken as variable).
8. PSNR between original and extracted watermark before and after attacks (taken as variable).

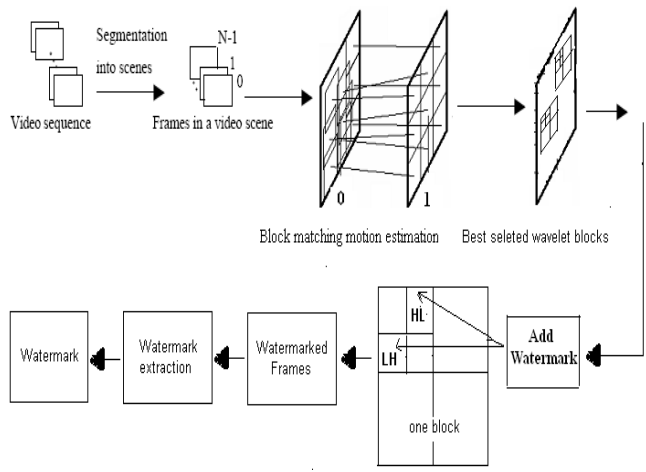

Fig. 5   Watermark Embedding and Extraction process in frequency domain

In this section are given the results of tests to assess the proposed system performance. The conducted tests have been directed to tune the involved parameters, such that the adopted values of the parameters are those which led to robust watermark embedding. Both embedding and extraction stages have the same parameters; each one plays an important role to achieve a high degree of robustness and imperceptibility. The presented results are from the tests conducted on different types of video samples like uncompressed ('Foreman') and compressed ('Spiderman') video sequences. The proposed motion estimator returns 225 motion blocks for compressed video while it returns 179 for uncompressed video for two successive frames. These numbers of block improve the condition that the number of selected best matched blocks must be greater than or equal to 128 blocks.

Table 1: Similarity against attacks LP: lowpass - HP:highpass filter

| Video types | Spatial Domain | | | | Frequency Domain | | | |
|---|---|---|---|---|---|---|---|---|
| | Before attacks | After attacks | | | Before attacks | After attacks | | |
| | | Adaptive quantization | LP filter | HP filter | | Adaptive quantization | LP filter | HP filter |
| Compressed 'Spiderman' | 24.41 | 5.34 | 14.3 | 12.03 | 43.58 | 12.32 | 30.97 | 27.35 |
| Uncompressed 'Foreman' | 27.49 | 6.62 | 17.99 | 16.43 | 37.92 | 8.89 | 20.08 | 18.91 |

Some tests are conducted when adaptive quantization and low or high filter processing are performed on watermarked frames. Since this scheme is developed in frequency domain, the similarity measure after any change is very close between the original and extracted watermark which lead this scheme to be very robust against any attacks. For the above reasons our scheme can be considered as a non blind robust technique. Table I and II present the comparison between the similarity and the PSNR results respectively in both spatial and frequency domain for three different types of attacks tested on compressed and uncompressed video sequence for two successive frames.

The results showed above indicate that the performance of the watermarking scheme in frequency domain is very high especially in the case when the adaptive quantization is performed on the motion blocks. In this case the quality of watermarked frame is considerably degraded after the attack (i.e., from 43.58 dB to 12.32 dB) while the similarity of the extracted watermark after the attack is remain close to the original watermark (i.e., from δ=1.0 to δ=0.6). In other hand, the effect of the frame filtering through low and high pass filters is presented to show the degree of robustness of the proposed scheme. Figure 5 shows the effect of different types of attack on the watermarked frame in the frequency domain.

The proposed watermarking schemes either in spatial or in frequency domain are tested when the size of watermark is taken by 512 bytes. The obtained results also give the robustness of the frequency domain scheme when the different types of attacks are performed.

Table 2: psnr (db) against attacks LP: lowpass - HP:highpass filter

| Video types | Spatial Domain | | | | Frequency Domain | | | |
|---|---|---|---|---|---|---|---|---|
| | Before attacks | After attacks | | | Before attacks | After attacks | | |
| | | Adaptive quantization | LP filter | HP filter | | Adaptive quantization | LP filter | HP filter |
| Compressed 'Spiderman' | 1 | 0.03 | 0.04 | 0.033 | 1 | 0.6 | 0.8 | 0.71 |
| Uncompressed 'Foreman' | 1 | 0.07 | 0.11 | 0.09 | 1 | 0.91 | 0.95 | 0.92 |

## 4. CONCLUTION

Digital watermarking provides more options and promises for multimedia security management. The solutions are more likely to remain application dependent and trade-offs between the conflicting requirements of low distortion, high capacity complexity, and robustness still have to be made. Before trustworthiness can be evaluated, possible attacks for specific applications have to be studied at the implementation stage.

This paper proposes a new video watermarking scheme based on motion estimation for color video sequence in a frequency domain. This technique is tested on compressed (taken from DVD high quality film) and uncompressed (taken by digital camera) video movies. The watermark is the random Gaussian distribution which is embedded into the motion regions between frames (HL, LH bands).

Experimental results show that the proposed new scheme has a higher degree of invisibility against the attack of frame dropping, adaptive quantization, and frame filtering than the previous developed scheme in spatial domain. The involved parameters are tuned in order to evaluate the performance of the algorithm by using the performance parameters (quality measure and similarity).

The future work will be the implementation of our scheme in different compressed video codec standards like (MPEG2, MPEG4). It can also be used for audio layer in video codec standards.

**Appendix**

*Terms and Definitions*

Motion estimation: A method for comparing two successive frames using one of the block matching techniques. This method helps to find out the motion regions between the frames where the watermark is embedded.

Robust watermarking: A method for embedding a secret message/watermark that is intended to be detectable even

49

after significant malicious or non-malicious manipulations on the host media where it is embedded.

Frequency based – watermarking: A method of embedding a watermark, a narrow-band signal, by spreading each bit of the watermark over several samples of the host media, a wide-band signal. The security resides in the secrecy of the Wavelet function.

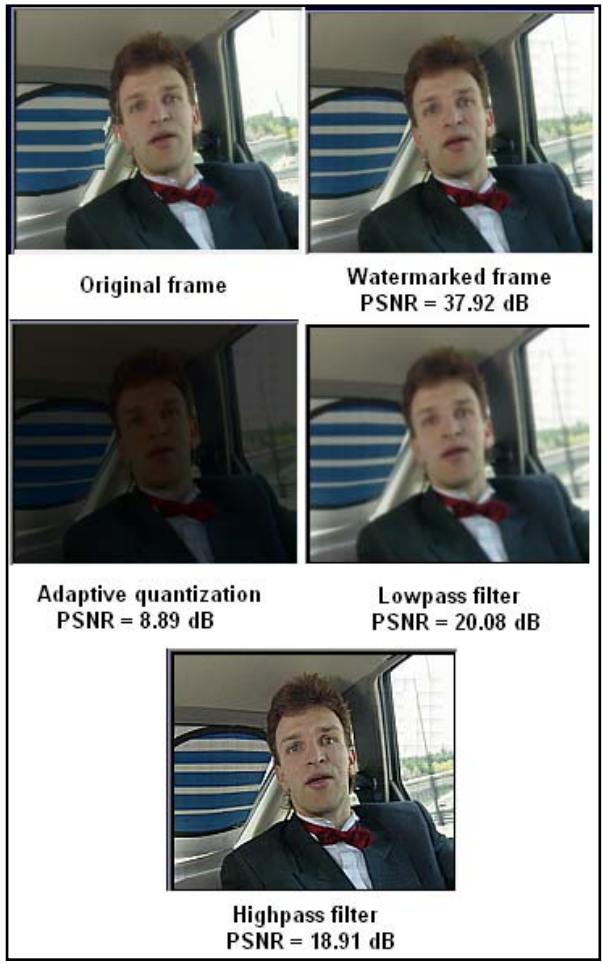

Fig. 5  PSNR against attacks in frequency domain

Multiband wavelet: The basic principle is the partitioning of the signal spectrum into several frequency bands, then embedding a watermark in each band separately.

Adaptive quantization: Given a set of watermarks and a set of quantizers with their adaptive quantization steps predefined according to the watermarks, quantization watermarking is a method of embedding a watermark in where a set of features are extracted from the host media and quantized to the nearest code of the quantizer corresponding to the watermark.

**Aree A. Mohammed** MSc in Atomic Physics 2001, MSc in Computer Science 2003, and PhD in Computer Science 2008. He is Head of Computer Department at College of Science / University of Sulaimani and member of its Scientific Committee. He has many research papers in Image Processing and Mutimedia fields either in International proceedings or in International Journals .

**Jamal A. Hussein** MSc in Computer Science 2007. Member of Scientific Committee in Computer Department at College of Science / University of Sulaimani. He has some research papers in the Watermarking field.